# Non-Paraxial Accelerating Beams

**Ido Kaminer, Rivka Bekenstein, Jonathan Nemirovsky, and Mordechai Segev**

Physics Department and Solid State Institute, Technion, Haifa 32000, Israel

**Abstract**

We present the spatially accelerating solutions of the Maxwell equations. Such non-paraxial beams accelerate in a circular trajectory, thus generalizing the concept of Airy beams. For both TE and TM polarizations, the beams exhibit shape-preserving bending with sub-wavelength features, and the Poynting vector of the main lobe displays a turn of more than 90°. We show that these accelerating beams are self-healing, analyze their properties, and compare to the paraxial Airy beams. Finally, we present the new family of periodic accelerating beams which can be constructed from our solutions.

The research on accelerating beams has been growing rapidly since it was brought into the domain of optics in 2007 [1]. An ideal paraxial accelerating beam is propagating along a parabolic trajectory, while preserving its amplitude structure indefinitely. The effect is caused by interference: the waves emitted from all points on the Airy beam profile maintain a propagation-invariant Airy structure, which shifts laterally along a parabola. This beautiful phenomenon has led to many intriguing ideas such as guiding particles along a curve [2], and to recent studies on shape-preserving accelerating beams in nonlinear optics [3,4]. In addition, it is possible to find beams accelerating along arbitrary curves, but such beams do not display shape-invariant propagation [5]. All of these beams are solutions of the paraxial wave equation, where the beam trajectory is fundamentally limited to small (paraxial) angles, and when it bends to larger angles - the beam is no longer shape-preserving. As a result, the transverse acceleration of Airy beams is always restricted to small angles. This restriction is a serious limitation, because spatial acceleration means that the propagation angle continuously increases, and eventually, after physically relevant distances, the beam trajectory inevitably reaches a steep angle, and the beam dynamics as a whole always goes into the non-paraxial regime. That is, a paraxial accelerating beam is moving along a curve which bends ever faster, and eventually it is bound to break its own domain of existence. Several attempts to find an accelerating beam beyond the paraxial regime have shown a complete breakup: part of the beam becomes evanescent while the other part quickly deforms while exhibiting only very small trajectory bending [6]. This brings about a series of fundamental questions: Can a beam accelerate at large non-paraxial angles? If it does, would such a non-paraxial accelerating beam be shape-preserving (as it is in the paraxial limit)? Such a beam should not be restricted by any physical parameters, and should be able to bend from a launch angle of zero all the way to angles close to 90° - vertical to the original direction of propagation. To put it exactly, the dynamics of light is governed by the Maxwell equations. Are there any accelerating shape-preserving solutions to the Maxwell equations?

Here, we present non-paraxial spatially accelerating shape-preserving beams. These accelerating beams are solutions to the full Maxwell equations, for time-harmonic linearly polarized fields. These non-paraxial accelerating beams propagate along a circular trajectory, therefore asymptotically reaching 90° angles, completing a quarter of a circle. As a validity test, we prove that taking these beams to the paraxial limit recovers the known paraxial Airy beams. This proves that, in the most general sense, an accelerating beam of EM waves (a wavepacket satisfying Maxwell's equations) is the beam we present here, and that the Airy beam found in [1] is simply the outcome of the paraxial approximation. We find the solutions for both TE and TM, where the beams exhibit shape-preserving bending with sub-wavelength features, and the Poynting vector of their main lobe displays a turn of more than 90°. We show that these accelerating beams are self-healing, and analyze their properties when they are emitted from finite apertures. Finally, we show that any given circular trajectory can support an entire family of accelerating solutions, whereby their superpositions form periodic accelerating beams.

We begin from Maxwell's equations in vacuum, for a TE-polarized electric field $\bar{E}=E_Y(x,z,t)\hat{y}$, obeying the Helmholtz equation

$$E_{xx} + E_{zz} - c^{-2}E_{tt} = 0. \tag{1}$$

Equation 1 has full symmetry between the *x* and *z* coordinates. Hence, it is logical to seek a shape-preserving beam whose trajectory resides on a circle. As initial condition, we use *E(x,z=0,t)* and let the beam propagate in the forward *+z* direction. Of course, such a beam cannot turn back to propagate in the $-z$ direction, hence the largest bending expected is a trajectory parallel to the *x*-direction. That is, the beam will asymptotically complete circular motion on a quarter of a circle. To seek such motion which is also shape invariant, it is convenient to transform Eq.1 to the rest frame of the beam. Since the motion is on a circle, we transform to polar coordinates *r*, *θ*, by taking $z = r \, sin(\theta)$, $x = r \, cos(\theta)$, and seek shape-preserving solutions of the form $E = U(r) \, e^{i\alpha\theta - i\omega t}$, where *α* is some real number, and *ω* is the temporal frequency. The result is a monochromatic beam, which is shape-preserving along any circular curve. The radial function *U(r)* must satisfy:

$$r\frac{d}{dr}\left(r\frac{dU}{dr}\right) + \left[-\alpha^2 + \left(\frac{\omega^2}{c^2}\right)r^2\right]U = 0. \qquad (2)$$

The exact solutions of Eq. 2 are the Bessel functions $U=J_\alpha[(\omega/c)r]$. [Actually, there is an additional family of solutions, also from the Bessel family, but those diverge at the origin, hence we will not discuss them here]. A related method was used recently by Hacyan [7], to find EM waves that accelerate relativistically in time.

To unravel the physics of our solution, it must be transformed back to the coordinates *x*, *z*; and separated into forward and backward propagating waves. We do it through the Fourier transform of the beam, which is confined to reside on a circle of radius $k=\omega/c=2\pi/\lambda$ in the $k_x$-$k_z$ plane (see the diagram in Fig.1). The top half (positive $k_z$) gives the forward propagating part of the beam, while the bottom half (negative $k_z$) give the backward propagating part. The forward propagating part is the actual forward-accelerating beam. Importantly, this beams does not have the Bessel structure, but only asymptotically half of it [8]. This can be calculated by integration over the top half of the circle (angles 0 to *π*), as sketched on the top half of the diagram on Fig.1.

$$E_Y^+(x,z) = \int_0^\pi e^{i\alpha k_\theta} e^{ik[x\cos(k_\theta)+z\sin(k_\theta)]} dk_\theta \equiv J_\alpha^+(kx,kz). \tag{3}$$

Where $J_\alpha^+$ is "half a Bessel" (because integrating the expression from $-\pi$ to $\pi$ yields the $\alpha$-order Bessel function $J_\alpha(kr)$). Here, $\alpha$ can be any real number (not necessarily integer) since we are not restricted to periodic boundary conditions (as the beam never completes a full circle). Figure 1a shows the accelerating solution of Eq. 3, for $\lambda=1\mu m$ and $\alpha=150$. The figure shows that the beam is indeed shape-preserving, but only up to an angle close to 90°. We should discuss the reason for this limit: Mathematically, it is clear that a Bessel solution is precisely shape-preserving. However, the physical beam is only "half a Bessel". In what sense is this beam shape-preserving? When $\alpha>0$, the exact Bessel beam is anti-symmetric at $z=0$ with respect to the origin: two main lobes positioned at opposite sides of $x=0$ and their oscillating tails stretch toward plus and minus infinity on their right and left sides respectively. But, the phase of the beam is what makes the beam anti-symmetric: it follows from the anti-clockwise rotation of the beam. This rotation makes the right half of the beam propagates forward and to the left (Fig.1a), while the left half propagates backward and to the right. The latter contradict the physical boundary conditions. When cutting the backward propagating waves in Fourier space, we are left with an almost-exact Bessel shape confined on the right side of the $x$ axis. Only when the bending gets close to 90°, the shape-preserving property breaks, where the two "half Bessel wavepacekts" were supposed to meet and interfere ([8]). Mathematically, larger $\alpha$ (larger angular momentum) gives better separation, hence also more accurate shape-preserving propagation.

The solution for the TM polarization is found through a similar procedure for the magnetic field, and from it the TM electric field components are found to be

$$E_Z^+(x,z) = \tfrac{1}{2} J_{\alpha+1}^+(kx,kz) + \tfrac{1}{2} J_{\alpha-1}^+(kx,kz)$$
$$E_X^+(x,z) = \tfrac{i}{2} J_{\alpha+1}^+(kx,kz) - \tfrac{i}{2} J_{\alpha-1}^+(kx,kz)$$
(4)

This TM solution is of special interest: each of its polarization components is not shape-preserving on its own, as shown in Figs. 1c and 1d, but the total intensity of the TM beam does preserve its shape. That is, as the beam bends by 90°, the power is transferred from the $x$-component to the $z$-component of the polarization. This shows that the accelerating beam not only bends but actually **rotates**, similar to the phase front of the beam which also rotates by 90°, staying normal to the beam trajectory at all times. A natural extension comes from the superposition of the TE and TM beams, which yields a vectorial solution of general polarization.

Generalizing Eq. 3 to arbitrary polarization gives the full family of vectorial 3D accelerating beams. We still choose the trajectory of acceleration in the $xz$ plane, but allow a plane wave in $y$. This leaves three functions in $k$ space that relate to the electric field via Eqs.5 (similar to Eq. 3)

$$E_j^+(x,y,z) = e^{ik_y y} \int_0^\pi f_j(k_\theta) e^{ik_{xz}[x\cos(k_\theta)+z\sin(k_\theta)]} dk_\theta \qquad j=x,y,z, \quad (5)$$

where $k_{xz}$, $k_y$ must satisfy the conditions: $k_{xz}\cos(k_\theta)f_x(k_\theta) + k_y f_y(k_\theta) + k_{xz}\sin(k_\theta)f_z(k_\theta) = 0$, and $k_{xz}^2 + k_y^2 = k^2$. Each polarization is therefore composed of a superposition of solutions of Eq.3 in the TE and the TM polarization, multiplied by a plane wave which only changes the effective wavenumber from $k$ to $k_{xz}$. Superpositions of fields with different $k_y$ should give beams which are confined in the y direction, extending the beams to 3D.

To highlight the impressive angle of bending, we note that it is actually possible to **double** the angle, by launching the beam at an angle opposite to the direction of bending. See Fig. 2a for an example of a beam that is launched at an angle of -65° and subsequently bends all the way to 65°, completing a turn of **130°**. In theory, the maximal bending is limited to asymptotically 180° because the boundary conditions allow only forward propagating waves. In practice, we measure the bending in Fig. 2a by the difference between the Poynting vectors of the main lobe at the incoming plane and at the outgoing plane. One can prove that the Poynting vector of the TM polarization is exactly the same.

Having found the accelerating solutions of Maxwell's equation, it is interesting to examine the small angles limit of the expression in Eq. 3, and see if it recovers the paraxial Airy solution. To do that, we recall a property of the Bessel function stating that the maximum of the main lobe occurs close to $x=\alpha/k$. Thus, to make the approximation in the correct range, we take $x= \alpha/k+\Delta x$ and assume $\Delta x$ and $z$ to be small. We also assume $\alpha$ to be very large, so that the exponent oscillates very fast and cancels out most of the contribution of the non-paraxial regime. In the limit of large $\alpha$, we expand the *cosine* and *sine* by a Taylor series around $\pi/2$, up to third order. The result is an integral that is solved analytically to yield

$$E_Y^+(x,z) \propto Ai\left(-2^{1/3}\alpha^{-1/3}k\Delta x - 2^{-2/3}\alpha^{-4/3}k^2z^2\right)e^{ikz+ikz\Delta x\alpha^{-1}-\frac{i}{3}k^3z^3\alpha^{-2}}, \qquad (6)$$

where $Ai$ is the Airy functions. Note the $z^3$ term characteristic of the paraxial Airy beam, indicating acceleration at a parabolic trajectory A direct consequence is that there is a ***unique relation*** between the parameter $\alpha$ and the acceleration (from the trajectory $\Delta x=-gz^2/2$), which is simply $g=k/\alpha$. Hence the acceleration $g$ is smaller for larger $\alpha$, which makes sense because higher orders of $\alpha$ give circular motion with larger radii, so the radial acceleration is indeed smaller. Another conclusion is that if we try to approximate an accelerating beam which is a superposition of several $\alpha$'s, we get a superposition of Airy beams with <u>different accelerations</u>.

This is why the paraxial accelerating beam must be a single Airy function, with a uniquely defined acceleration, whereas the non-paraxial accelerating beams can support a family of beams of different shape (different α's) which all accelerate at the same trajectory. Finally, for small values of $α$, we find accelerating beams that <u>cannot</u> exist in the paraxial regime at all. Those beams are mainly made up of very high spatial frequencies, hence trying to construct an Airy beam with these constituents gives rise to a beam that breaks up after a very short propagation distance. See Fig. 1b for an example, with $α$=150, where the parabolic acceleration survives for a very short distance only.

It is now worth while to compare the known features of the Airy beam to our new non-paraxial accelerating beam. To this end, we notice that the self-healing effect [9] also exists in our non-paraxial beams: see Fig. 2b for an example, where the main lobe is initially cut out (blocked). The second lobe gets more power and its trajectory bends more – to replace the first lobe; see the dotted black line on Fig. 2b marking the trajectory of the original first lobe. When this "replacement" occurs, each lobe is shifted by a steeper bending to replace the lobe on its left. Another property common for both the Airy beam and the non-paraxial accelerating beam is that both are not square integrable, hence they carry infinite power. In this context, launching either of them from a finite aperture yields a beam that accelerates over a finite propagation range only. As with the Airy beam, longer tail in the non-paraxial case allows for more lobes to exhibit shape-preserving propagation for larger distances. Interestingly, the reason that the non-paraxial accelerating beam carries infinite power comes from only two singular points in $k$-space, which are placed at the edges (-$k$ and $k$) bordering the regime of evanescent waves. Removing these singular points leaves an accelerating beam of finite power, since then the $k$-space spectrum becomes square intengrable (unlike the paraxial spectrum of the Airy which is unbounded). Physically, some part of the spatial spectrum will always be removed, since the edges of the *k*-space represent waves of zero velocity. At the same time, a rather short tail (of about twice the radius of the trajectory) is sufficient to make the first lobes to bend into a deep non-paraxial angle (more than 50°). Consequently, a non-paraxial accelerating beam launched from a finite aperture (thus carrying finite power) will bend on a circular curve while maintaining a virtually propagation-invariant shape for the majority of the physically-accessible one quarter of a circle. Finally, another difference between the Airy beam and the non-paraxial accelerating beam is that, unlike the Airy beam, the Bessel-like non-paraxial accelerating beams cannot be simply scaled (by squeezing or stretching the *x* axis) to control the acceleration curve. Rather, in the non-paraxial case different values of *α* imply different orders of the Bessel-like function, which affects the lobes widths indirectly.

When coming to examine the non-paraxial accelerating solutions of Eqs. 3 and 4, we note that any superposition of these solutions, with different values of *α*, also gives an accelerating beam which is propagating on the same curved trajectory. Such superposition accelerates in unison, but it is not shape preserving: it is a breather, with a periodicity depending on the difference between the values of *α*. Thus, an infinite family of **periodic accelerating beams** can be generated from superpositions. Figure 2c displays such a periodic accelerating beam. Note that some of the periodic accelerating beams have finite power, due to the destructive interference of the tails. Mathematically, this happens when the singular points in *k*-space ($k_\theta=0,\pi$) are canceled by summing two or more waves, as in the case of the *x*-component of the TM polarization (which is also a legitimate periodic solution for TE). Many other examples of **finite power** periodic accelerating beams can be generated from Eq.5 for mixed polarizations.

Before closing, we note that accelerating beams can be also found through methods relying on caustics [5,10,11]. One can actually use the caustics method to design phase masks for launching beams that accelerate along any convex trajectory [5,10]. Interestingly, a recent paper [10] uses the caustic method to generate non-paraxial accelerating beams. This method is based on ray-optics logic but taken into non-paraxial angles. This way, accelerating beams moving along an arbitrary curve can reach large bending angles. However, while this method constrains the main lobe to accelerate along the predesigned curve, it does not determine how the rest of the beam is propagating. In practice, the beam is considered accelerating but it is not shape-preserving: after some distance, diffraction effects smear the beam structure and acceleration stops [5,10]. In this respect, such "caustic-designed accelerating beams" are different from the paraxial Airy beams, from nonlinear accelerating beams [3], and even from the non-paraxial accelerating beams described here, which are all shape-preserving: the entire beam is accelerating with a propagation-invariant amplitude, whereas caustic-designed accelerating beams are not meant to be propagation-invariant.

To summarize, we have found non-paraxial accelerating beams and non-paraxial periodically-oscillating accelerating beams. These beams are the full vector solutions of Maxwell's equation for accelerating beams. [the same as in the abstract. I think we should claim the "the"] Taking these beams to the paraxial domain recovers the known Airy beams. Interestingly, this work has unexpected implications in optics and beyond. First, it is now clear that the phenomenon of accelerating waves is not merely the result of a specific unusual behavior of the Schrödinger equation (which is equivalent to the paraxial wave equation), as one may think from reading the first paper on this subject [12]. Rather, it is a **fundamental property of all electromagnetic fields, because we have found accelerating solutions of the full Maxwell equations**. Second, **non-diffracting** beams are no longer necessarily Bessel beams which propagate on a straight line, but now include also self-bending beams. To complete the picture, future work should study the possibility of 3D accelerating beams of Maxwell's equations, including those with trajectories that do not lie in a single plane. In practical terms, this work brings accelerating optics into the sub-wavelength regime, through the less-than-wavelength features of our solutions, facilitating higher resolution for particle manipulation.

---

References (without titles)

Figure 1

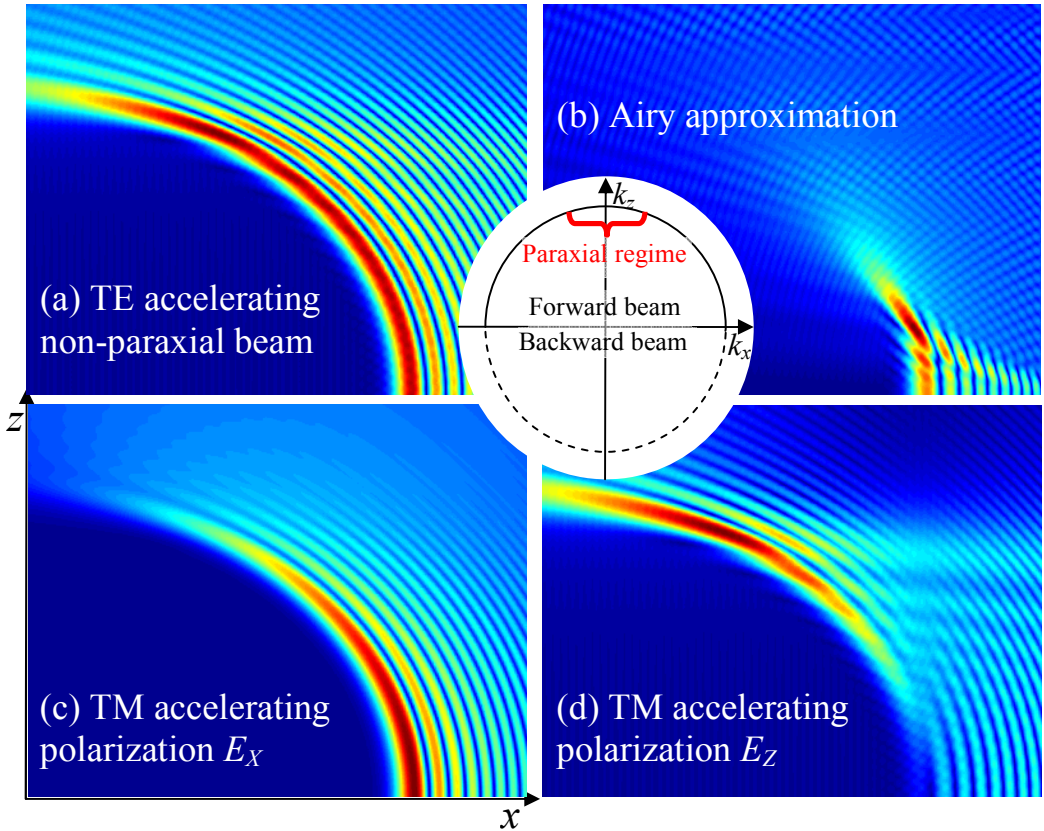

Accelerating beams of the Maxwell's equations. [a] The forward accelerating beam of TE polarization ($\alpha$=150) reaches a trajectory which is almost vertical, while exhibiting shape-preserving acceleration. [b] The paraxial approximation yields an Airy beam which accelerates only for a short distance before breaking up. [c,d] The forward propagating beam of TM polarization ($\alpha$=150). The power transfers from the $x$ polarization (c) to the $z$ polarization (d). All figures are simulated with $\lambda$=1μm and in a square of 35μm × 35μm. The diagram shows the Fourier plane where the beam is confined in a circle describing the propagating plane-waves. The top (bottom) half stands for the forward (backward) beam.

Figure 2

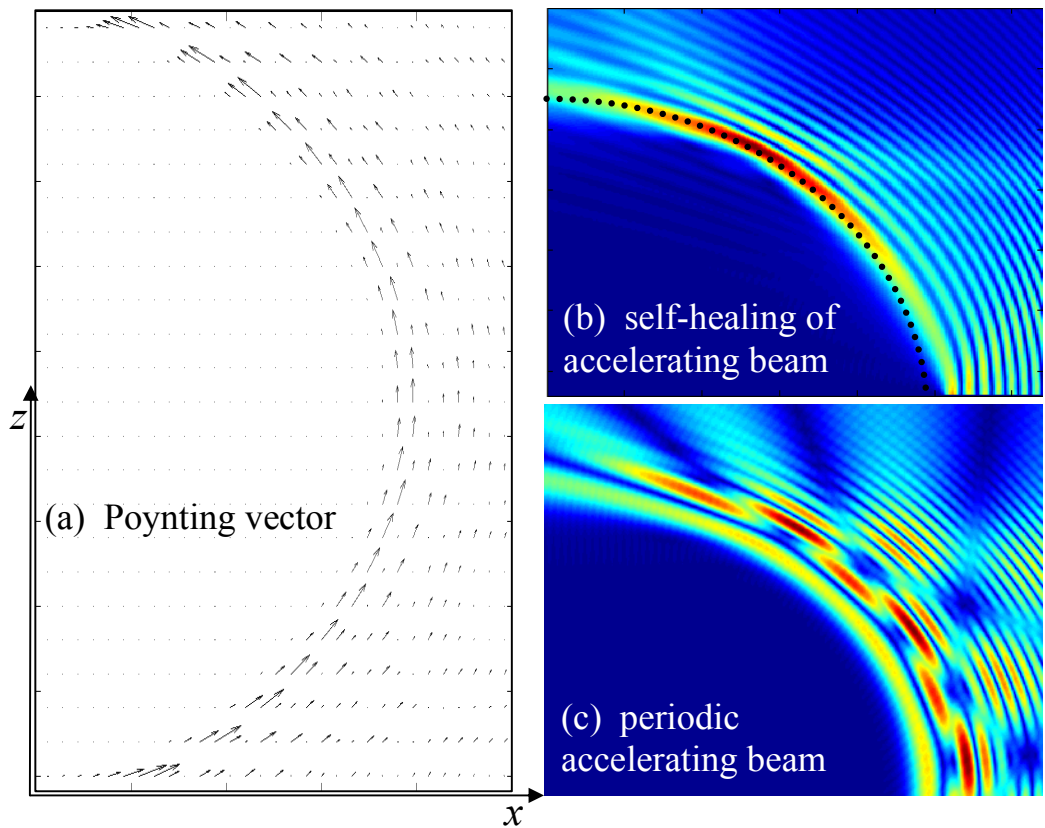

Properties of the Bessel-like accelerating beams. [a] The Poynting vector of the TE/TM polarization, when launched with an initial angle and bends by 130°. [b] The self-healing property "reviving" the first lobe that was initially cut out. The dashed-black curve describes the original trajectory of the first lobe of a perfect accelerating beam. [c] A periodic accelerating beam composed of two jointly-accelerating components $\alpha$=150 and $\alpha$=165 with equal coefficients. All figures are simulated with $\alpha$=150, $\lambda$=1μm and in a square of 30μm × 45μm (a) or 35μm × 35μm (b,c).